\documentclass[aoas, preprint]{imsart}

\RequirePackage[OT1]{fontenc}
\RequirePackage{amsthm,amsmath}
\RequirePackage[longnamesfirst]{natbib}
\RequirePackage[colorlinks,citecolor=blue,urlcolor=blue]{hyperref}

\usepackage{graphicx}

\arxiv{arXiv:1410.6059}

\startlocaldefs
\numberwithin{equation}{section}
\theoremstyle{plain}

\endlocaldefs

\begin{document}

\begin{frontmatter}
\title{Integer percentages as electoral falsification fingerprints}
\runtitle{Integer percentages as falsification fingerprints}

\begin{aug}
\author{\fnms{Dmitry} \snm{Kobak}\thanksref{m1}\ead[label=e1]{dmitry.kobak@neuro.fchampalimaud.org}},
\author{\fnms{Sergey} \snm{Shpilkin}\ead[label=e2]{podmoskovnik@gmail.com}}
\and
\author{\fnms{Maxim} \fnms{S.} \snm{Pshenichnikov}\ead[label=e3]{maxim.pshenichnikov@gmail.com}}

\runauthor{D. Kobak et al.}

\affiliation{Champalimaud Centre for the Unknown\thanksmark{m1}}

\address{Dmitry Kobak\\
Champalimaud Centre for the Unknown,\\Lisbon, Portugal\\
\printead{e1}}

\address{Sergey Shpilkin\\
Moscow, Russia\\
\printead{e2}}

\address{Maxim Pshenichnikov\\
Groningen, Netherlands\\
\printead{e3}}

\end{aug}

\begin{abstract}
We hypothesize that if election results
are manipulated or forged, then, due to the well-known human attraction to round numbers,
the frequency of reported round percentages can be increased.
To test this hypothesis, we analyzed raw data
from seven federal elections held in the Russian Federation during
the period from 2000 to 2012 and found that in all elections since
2004 the number of polling stations reporting turnout and/or leader's
result expressed by an integer percentage (as opposed to a fractional
value) was much higher than expected by pure chance. Monte Carlo simulations
confirmed high statistical significance of the observed phenomenon
thereby suggesting its man-made nature. Geographical analysis showed
that these anomalies were concentrated in a specific subset of Russian
regions which strongly suggests its orchestrated origin. Unlike previously
proposed statistical indicators of alleged electoral falsifications,
our observations can hardly be explained differently but by a widespread
election fraud.
\end{abstract}

\begin{keyword}[class=MSC]
\kwd[Primary ]{62P25}
\kwd[; secondary ]{91F10}
\end{keyword}

\begin{keyword}
\kwd{electoral falsifications}
\kwd{Russian elections}
\end{keyword}

\end{frontmatter}

\section{Introduction}


Human attraction to round numbers (such as e.g. multiples of 5 or
10) is a well-known psychological phenomenon, frequently observed
e.g. in sports, examinations \citep{pope2011}, stock markets \citep{harris1991, kandel2001, osler2003}, pricing \citep{klumpp2005}, tipping \citep{lynn2013}, census
data \citep{yule1927}, survey results \citep{crawford2014sex}, etc. Excess of round numbers in such data is sometimes called ``heaping'' \citep{crawford2014sex}. One likely interpretation of this phenomenon is that round numbers act as
reference points when people are judging possible outcomes \citep{pope2011}. Recently,
this phenomenon has helped catching data manipulations or forgery
in cases of scientific misconduct \citep{simonsohn2013}. Here we hypothesize that a similar
effect could show up in electoral data as well: if election results
are manipulated or forged, then the frequency of reported round percentages
should be increased. To test this hypothesis, we analyzed raw data
from seven federal elections held in the Russian Federation during
the period from 2000 to 2012 and compared it to similar elections in other countries.

Russia presents an unusual case of a country where all raw electoral data
are freely available for inspection, but election results are allegedly
subject to forgery. Indeed, Russian federal elections after the year
2000 have often been accused of numerous falsifications, in particular
on the grounds of multiple anomalies in the raw election data \citep{mikhailov2004, mebane2006, myagkov2009, mebane2010, klimek2012, simpser2013, ziegler2013, enikolopov2013}.
Convincing as these indictments are, they all have serious limitations:
some are indirect \citep{mikhailov2004, myagkov2009} or model-based \citep{klimek2012}, while
the reported anomalies can in principle be explained by social, geographical
or other confounding factors \citep{coleman2004, churov2008, hansford2010}. Some are based on
field experiments \citep{enikolopov2013} conducted in one single city; some
rely on Benford's law \citep{mebane2006, mebane2010}, were criticized for that
\citep{deckert2011}, and are now deemed inconclusive \citep{mebane2013, mebane2013using, mack2013}.
The position of Russian authorities has always been that the official
results of all Russian elections are genuine\footnote{Press conference of Vladimir Putin, 2011: \url{http://www.rg.ru/2011/12/15/stenogramma.html} (in
Russian); Interview with the press attache for the president, Dmitry Peskov, 2011: \url{http://lenta.ru/
news/2011/12/12/noeffect} (in Russian).}.

Here we focus on another statistical anomaly: elevated frequency of round percentages in the election results. Anomalously high incidence of multiple-of-five percentages in some Russian federal elections has been observed before by one of us \citep[as reported in][p 201]{buzin2008crime}, used in our preliminary work \citep{kobak2012}, and also mentioned by Mebane et al. \citep{mebane2009comparative, mebane2014geography, kalinin2010understanding, mebane2013using}. Here we demonstrate that it is only a part of a more general phenomenon: anomalously high incidence of high integer percentages. We used Monte Carlo simulations to confirm statistical significance of this anomaly and measure its size. We argue that it presents a convincing evidence of election fraud that was absent in 2000 and 2003 federal elections, appeared in 2004 and has remained ever since.

\section{Materials and methods}


\subsection{Background}

Our analysis involves seven Russian federal elections: four presidential
(2000, 2004, 2008, and 2012) and three legislative ones (2003, 2007,
2011). In each of these elections, the winner was either Vladimir
Putin (2000, 2004, 2012) or his prot\'{e}g\'{e} Dmitry Medvedev (2008), or
the pro-government party United Russia (2003, 2007, 2011). We always
refer to the winner candidate or party as ``leader''. 

The legislative elections in 2007 and 2011 were conducted under a
nationwide proportional system (i.e. the seats in the parliament were distributed between parties according to the proportion of votes for each party). The 2003 legislative election was
mixed, with half of the deputies elected in a nationwide proportional
election and another half in majoritarian districts (with each district electing one member of parliament); here we consider
only the proportional part. The presidential elections are direct (i.e. people vote directly for the candidates and not for the electors as is the case in indirect elections),
and in all elections under consideration the winner was determined
in the first round, although the second round was in principle possible. 

The total number of registered voters in Russia in 2000--2012 was
about 108 million (107.2 to 109.8 million for different elections),
and the total number of polling stations varied from $95\,181$ to
$96\,612$. At a lower level, the polling stations are
grouped into constituencies (2744 to 2755 in total) corresponding
to administrative territorial division. Constituencies vary in size
and may contain from a few to more than a hundred polling stations.
Constituency-level electoral commissions gather voting data in the
form of paper protocols from the polling stations and enter them into
the nationwide computerized database (``GAS Vybory''). 

At a higher level, in 2012 Russia was divided into 83 federal regions.
The number of regions slightly decreased from 2000 to 2012, as several
regions were merged. In our analysis of earlier elections we combined
the regions that would later be merged officially to keep consistency
with the 2012 nomenclature.

\subsection{Data}
\label{sec:data}

The raw election data with detalization to polling stations are officially
published at the website of Russian Central Election Committee (\textit{izbirkom.ru})
as multiple separate HTML pages and Excel reports. For our analysis,
these data were downloaded with custom software scripts to form a
joint database. The accuracy of the resulting databases was verified
by checking regional subtotals and comparing a number of randomly
chosen polling stations with the respective information at the official
website. The parts of election databases relevant for the current study are provided as Supplementary Materials.

For the 2003--2012 elections, detailed data are available for each and
every polling station in the country. For the 2000 election, the polling
station level data are missing for the Republics of Chechnya and Sakha-Yakutia,
and for several constituencies in other regions; available data cover
$91\,333$ polling stations ($\sim$95\% of total number) and 105.6
million voters (97.3\% of total number).

For each polling station $i$ the following values (among many more)
are available: the number $V_{i}$ of registered voters, the number
$G_{i}$ of given ballots\footnote{This is the sum of ballots given to the voters at the
polling station at the election day, ballots given to the voters
outside of the polling station at the election day (in Russia it is
possible to vote at home), and ballots given during early voting.}, the number $B_{i}$ of cast ballots (sum
of valid and invalid ballots), and the number $L_{i}$ of ballots
cast for the leader. In some cases, $B_{i}$ is not equal to $G_{i}$
due to taken away (not cast into the box) ballots; this fraction is
small, $\sim$0.1--0.3\%. According to Russian electoral laws\footnote{Federal law regulating parliamentary elections: \url{http://cikrf.ru/law/federal\_law/zakon\_51/gl11.html}.},
turnout $T_{i}$ at a given polling station is defined as $T_{i}=G_{i}/V_{i}\cdot100$\%
and leader's result $R_{i}$ as $R_{i}=L_{i}/B_{i}\cdot100$\%. Although
turnout lost its legal significance after 2006 electoral law
amendments that abolished turnout thresholds, \textit{de facto} it is still
customarily included in high-level official reports and, as our analysis
will show, remains an important reporting figure at lower levels of the
electoral system.

In special cases where $V_{i}$ is not defined beforehand
(e.g. in temporary polling stations located at the train stations
or airports), $V_{i}$ is officially reported as equal to $G_{i}$,
automatically resulting in 100\% turnout (3--5\% of all stations). We exclude all such stations from our analysis.

Official election results are reported at the national level only
and are calculated as $\sum G_{i}/\sum V_{i}\cdot100$\% and $\sum L_{i}/\sum B_{i}\cdot100$\%
for turnout and result, respectively. Although the results at lower
levels (region, constituency, polling station) do not have any legal
significance, they are nevertheless available at the Central Election
Committee official website down to single polling station level.

\subsection{Data from other countries}

We used election data from three countries besides Russia: 2011 general election in Spain, 2010 presidential election in Poland (1st round), and 2009 federal election in Germany (Zweitstimmen, i.e. party votes). These three elections were chosen because the data are publicly available down to the single polling station level, and because the number and size of polling stations are comparable to those in Russia. The winners of these elections were the People's Party, Bronis{\l}aw Komorowski, and the CDU/CSU coalition respectively.

The Polish dataset is directly available at the official website  in CSV format (\textit{prezydent2010.pkw.gov.pl}), and the dataset for Spain is provided at the official website (\textit{www.infoelectoral.mir.es}) in a custom format that requires decoding. The German dataset in CSV format was obtained by post on a CD after a request to the German federal returning officer (\textit{bundeswahlleiter.de}).

In all cases $V_{i}$ is the number of registered voters and $L_{i}$ is the number of ballots cast for the leader. In Spain $G_{i}$ is defined as the sum of invalid, empty, and valid cast ballots and $B_{i}$ as the sum of empty and valid cast ballots. In Germany $G_{i}$ is defined as the sum of invalid and valid cast ballots and $B_{i}$ as the number of valid ballots. In Poland $G_{i}$ is defined as the number of given ballots and $B_{i}$ as the number of valid cast ballots.

The total number of polling stations is $25\,774$ for Poland, $59\,928$ for Spain, and $75\,096$ for Germany (in Germany we excluded from the total number of $88\,705$ stations those $9609$ of them lacking information about the number of registered voters).

\section{Results}

\subsection{Integer anomaly}

To avoid any a priori assumptions about what constitutes a ``round''
percentage (multiple of 10? multiple of 5? any even number?), we chose to look at all
integer percentages. As it is often impossible to achieve an exactly
integer percentage at a given polling station because voter and ballot
counts are integer (e.g. on a polling station with 974 registered
voters the closest possible value to 70\% turnout is 70.02\% with
682 people participating in the election), we counted as integer
all percentage values deviating from an integer by at most 0.05 percentage
points. With characteristic number of ballots per station being $\sim$1000,
such precision could almost always be achieved.

\begin{figure}[t]
\centering
\includegraphics[width=1\linewidth]{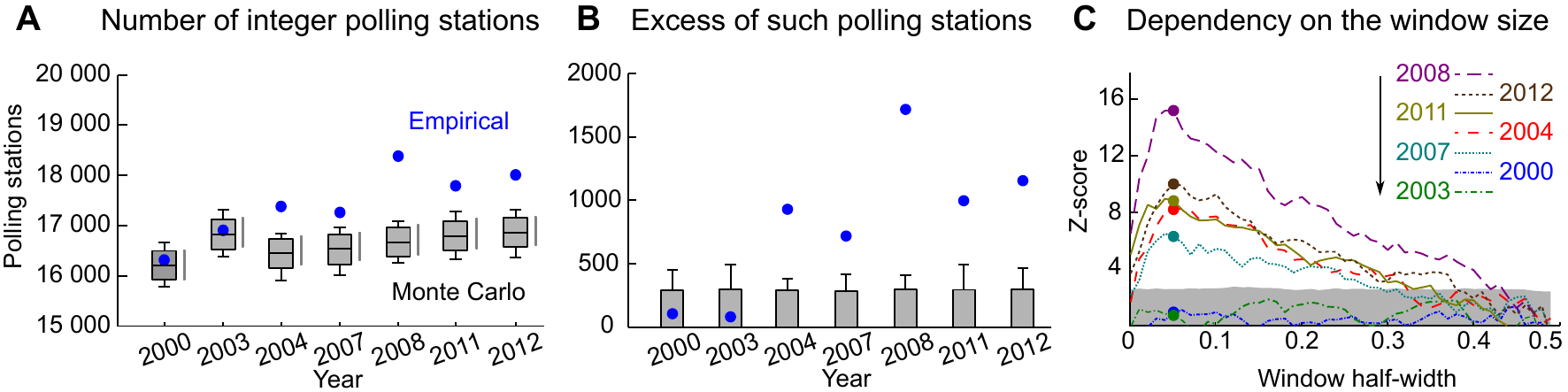}
\caption{\textbf{(A)} Number of polling stations with integer turnout or result
percentage value, $\pm 0.05\%$ (blue dots). Box plots show distributions of the
same quantity expected by chance, obtained from binomial Monte Carlo simulations.
Boxes show 0.5\% and 99.5\% percentiles together with the median value
(horizontal line), whiskers extend from the minimal to the maximal
value obtained in all 10\,000 Monte Carlo runs. Grey vertical lines next to the box plots show 99\% percentile intervals of beta-binomial Monte Carlo simulations. \textbf{(B)} The
same as in (A), but with mean value of Monte Carlo distribution subtracted
from empirical values for each year to highlight the deviations between the two. \textbf{(C)} The same as in (B), but computed with various window sizes around integer values and converted to $z$-scores: empirical value minus mean Monte Carlo value, divided by the standard deviation of Monte Carlo values. Each curve corresponds to one particular year (see legend). Grey shading shows 99.5\% Monte Carlo percentiles ($z\approx 2.5$).}
\label{fig:anomaly}
\end{figure}

For each year we counted the number of polling stations where either
turnout or leader's result were given by an integer percentage $\pm 0.05\%$ (Figure~\ref{fig:anomaly}A,
dots); we will call those ``integer polling stations''. Prior to this counting, we excluded all polling stations with turnout or result being over 99\% because a large number of stations are reported with a formal turnout of 100\% which is an integer; we wish to exclude these from the analysis (see Section~\ref{sec:data}). All polling stations with less than 100 registered voters were excluded as well because those are often temporary polling stations with some special status. The number $q$ of integer polling stations among the remaining $n$ stations is our main statistic.

One could think that in a fair election the chance for the turnout to be given by an integer$\pm 0.05\%$ is $1/10$, and the same is true for the leader's result; it follows that $q$ should be approximately equal to $[1-(9/10)^2]n=0.19n$. However, the distribution of $q$ is affected by the distribution of polling station sizes: e.g. at a polling station with 100 registered people, \textit{all} possible turnouts are integer. In particular for small polling stations the probability of $q/n$ can noticeably deviate from 0.19. For that reason we used Monte Carlo simulation to sample from the null distribution of $q$.

Specifically, Monte Carlo simulations were based on the following null hypothesis:
first, the election outcome at each polling station represents the
true average intentions of voters at that particular location, and
second, each person at each polling station votes freely and independently.
Accordingly, for each polling station $i$ we modeled the turnout as a random variable $$T^\mathrm{MC}_i = G^\mathrm{MC}_i / V_i \cdot 100\%, \;\;\; G^\mathrm{MC}_i \sim \mathsf{Binom}(V_i, G_i/V_i),$$ and the leader's result as a random variable $$R^\mathrm{MC}_i = L^\mathrm{MC}_i / B_i \cdot 100\%, \;\;\; L^\mathrm{MC}_i \sim \mathsf{Binom}(B_i, L_i/B_i).$$ Note that for
large $V_{i}\gg 1$ and $B_{i}\gg 1$ this yields the following Gaussian approximation (not used in actual simulations):
\begin{align*}
T^\mathrm{MC}_i &\mathrel{\dot\sim} \mathcal N(T_i, \,\,T_i(100-T_i)\,/\,V_i) \\
R^\mathrm{MC}_i &\mathrel{\dot\sim} \mathcal N(R_i, \,R_i(100-R_i)/B_i),
\end{align*}
meaning that e.g. for a polling station with 1000 registered voters and 60\% turnout $T^\mathrm{MC}\mathrel{\dot\sim}\mathcal N(60, 2.4)$.

After generating $T^\mathrm{MC}_i$ and $R^\mathrm{MC}_i$ for all $i\in\{1, \dots, n\}$, the main statistic $q$ was computed as described above, and this procedure was repeated $10\,000$ times to obtain $10\,000$ values of $q$ sampled from the null distribution (a typical run of $10\,000$ Monte Carlo iterations took $\sim$8h on a single core of an Intel i7 3.2 GHz processor). As a result, for each year we obtained a distribution of the amount of
integer polling stations that could have arisen purely by chance,
under the null hypothesis of no manipulations (Figure~\ref{fig:anomaly}A, box plots). 

Figure~\ref{fig:anomaly}A shows that in 2000 and 2003 the empirical number of integer
polling stations (computed from the actual electoral data) falls well
within the 99\% percentile interval of the Monte Carlo values. However,
starting from 2004, the empirical number by far exceeds all $10\,000$
Monte Carlo values, meaning that the observed number of integer polling
stations could almost certainly not have occurred by chance. Therefore, the
null hypothesis of no manipulations of the electoral results can be
rejected with $p<0.0001$. Figure~\ref{fig:anomaly}B shows how much the number of integer polling
stations in each year exceeded the mean Monte Carlo value (i. e. the
most likely value in the absence of manipulations): starting with
2004, the resulting ``anomaly''
is around 1000 polling stations, and it peaks in 2008 reaching
almost 2000 polling stations.

The exact size of the anomaly depends on the window size used to define what percentage values are counted as being close enough to an integer. However, the anomaly sizes remain almost the same with windows ranging from around $\pm 0.05 \%$ size (used above) to around $\pm 0.15 \%$, and the $z$-scores peak around $\pm 0.05\%$ as shown on Figure~\ref{fig:anomaly}C. Larger windows yield smaller and less significant anomalies (see also Figure~\ref{fig:meanpeak} below), dropping to zero at $\pm 0.5 \%$ window that simply counts \textit{all} polling stations and therefore yields $q=n$.

\subsection{Controls}

We ran a number of controls to ensure that the integer anomaly is a real and 	nontrivial effect.

\begin{figure}[t]
\includegraphics[width=1\textwidth]{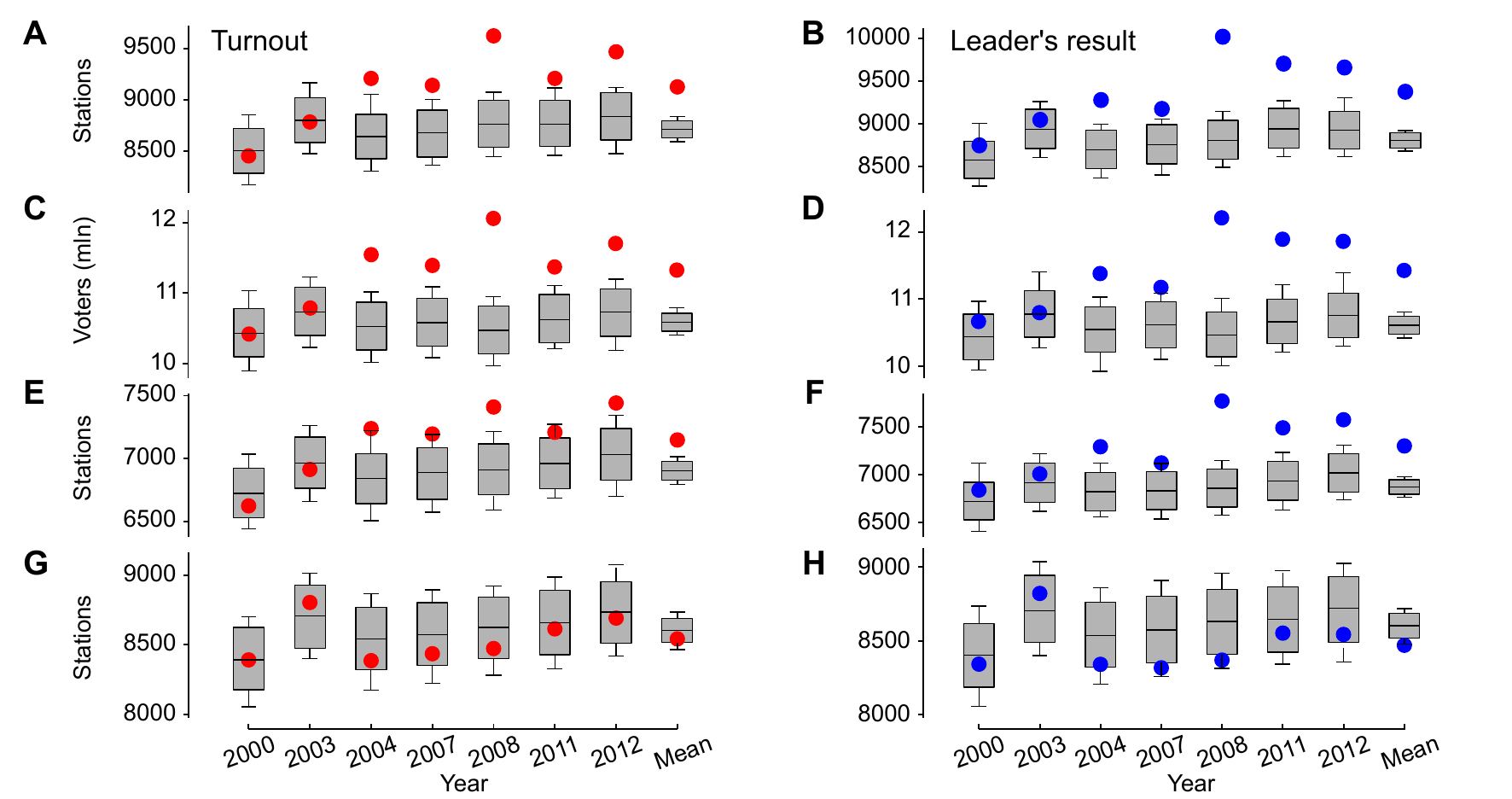}

\caption{\textbf{(A--B)} The same as in Figure~\ref{fig:anomaly}A, but the number of integer
polling stations was computed separately for turnout (A) and leader's
result (B). Here and below, in addition to each year, we show the mean value over
all seven years, in order to increase signal-to-noise ratio. \textbf{(C--D)}
Number of registered voters on all integer polling stations. 
\textbf{(E--F)} Number of integer polling stations, after excluding
all polling stations where either the number of given ballots or the
number of registered voters ended in zero (for turnout, E); or where
either the number of cast ballots or the number of ballots cast for
the leader ended in zero (for leader's result, F).
\textbf{(G--H)} Number of half-integer polling stations, i.e. polling
stations reporting turnout (G) or result (H) differing by at most
0.05 percentage points from a half-integer percentage.}
\label{fig:controls}
\end{figure}

First, the same anomaly can be computed for turnout and leader's result separately. In both cases, the number of integer polling stations is well above the whole Monte Carlo range, as before (Figure~\ref{fig:controls}A--B).

Second, one can worry that high number of integer polling stations can arise in fair elections due to artifacts of division of small integers \citep{johnston1995} (even though small polling stations with less than 100 registered voters were excluded from our analysis). This cannot be so, because the same artifacts would then also appear in Monte Carlo simulations and would not make the empirical number of integer polling stations appear exceptional. Still, in addition to counting integer polling stations, we also computed the sum of registered voters at all integer polling stations (Figure~\ref{fig:controls}C--D). This metric is mostly influenced by large polling stations. Significant and substantial anomalies in all years after 2004 confirm our conclusions and indicate that the integer anomaly is not an effect of small stations.

Third, it has been proposed before \citep{beber2012} that a higher than expected
number of polling stations reporting round (i.e. ending in 0) counts
of registered voters, cast ballots, etc., can be taken as an evidence
of fraud. Nonetheless, one can argue that such round counts can occur
due to ``innocent'' (but still
illegal) rounding in the exhausting manual ballot counting and do
not necessarily imply a malicious fraud. Crucially, this is not the
case for the anomaly reported here because the official precinct paper
protocols in Russia contain only ballot counts, and do not contain
either turnout or leader's result in percent. However, the performance
of a ballot station is likely to be judged by higher authorities by
the shown percentages, prompting to fiddle with the ballot counts
until appealing \textit{percentages} are obtained. Notably, in most cases this
requires non-round ballot counts. We have checked consistency of this
argument by excluding all polling stations with round counts: the number of integer-turnout polling stations was computed without counting stations where either $T_i$ or $V_i$ ended on zero, and the number of integer-result polling stations was computed without counting stations where either $R_i$ or $B_i$ ended on zero (Figure~\ref{fig:controls}E--F). This decreased the anomalous number of round percentages only slightly.

Fourth, we computed the number of polling stations with turnout or result differing by at most $\pm 0.05\%$ percentage points from a half-integer (as opposed to integer) percentage. This serves as a consistency check that, as expected, shows no significant effect in any year (Figure~\ref{fig:controls}G--H).

Fifth, do our conclusions depend on the particular details of the Monte Carlo simulation? We argue that they do not. In addition to the binomial distribution, we also used the beta-binomial one:
\begin{align*}
&G^\mathrm{MC}_i \sim \mathsf{Binom}(V_i, p_i),\;\;\; p_i \sim \mathsf{Beta}(G_i+1, V_i-G_i+1),\\
&L^\mathrm{MC}_i \sim \mathsf{Binom}(B_i, p_i),\;\;\; p_i \sim \mathsf{Beta}(L_i+1, B_i-L_i+1).
\end{align*}
This choice is motivated as follows. The observed turnout and leader's result are not
exact measurements of voters' intentions, and one can estimate the
conditional distribution of true voters' intentions $p_i$ given the observed
value and the uniform prior --- this leads to the beta distribution. When a beta-distributed $p_i$
is used as a parameter for the binomial distribution, the compound
distribution becomes beta-binomial. We performed beta-binomial Monte Carlo simulations (1000 iterations), and the null distributions hardly changed at all (Figure~\ref{fig:anomaly}A).

Binomial distribution has been successfully used to describe statistics
of election results across very different countries \citep{borghesi2012}.
For some countries, the data suggest that voters tend to vote in clusters, corresponding e.g. to families. This positive correlation between voters leads to higher variance of simulated outcomes
at each polling station compared to the binomial distribution,
and the integer percentages observed in the actual data would be smeared
even stronger in the simulated data. To confirm this, we ran Monte
Carlo simulations with various values of cluster sizes up to 10  and did not observe any excess of integer polling stations in the simulations.

The behaviour of actual voters is likely described by even more complex distributions, capturing perhaps some correlations between candidates and non-independence of voters. The existing evidence suggests that the more realistic distributions are overdispersed as compared to the binomial one \citep{borghesi2012}. Whereas underdispersed distributions are theoretically possible, they seem unlikely to occur in real life (to yield noticeable underdispersion, voters should e.g. be precisely orchestrated or should \textit{en masse} exhibit strong negative correlations). For these reasons we believe that the binomial assumption is conservative for the current purposes. 

\begin{figure}[h]
\includegraphics[width=0.5\textwidth]{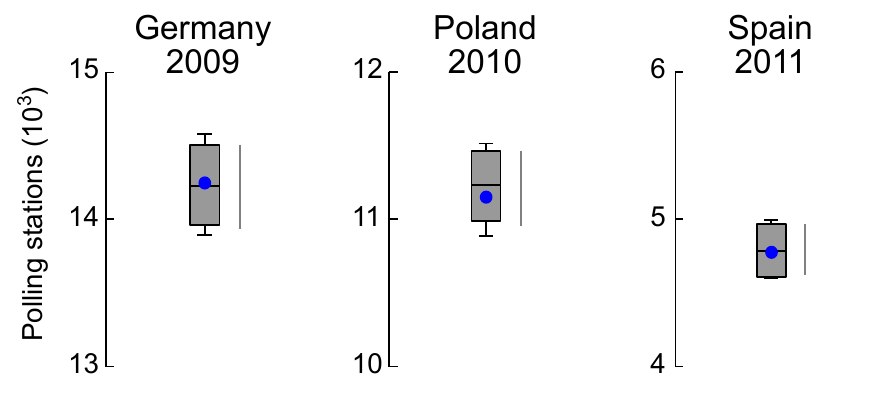}
\caption{Number of polling stations with integer turnout or result
percentage value in three different elections outside of Russia (blue dots). Box plots show distributions of the
same quantity expected by chance, obtained from binomial Monte Carlo simulations (as in Figure~\ref{fig:anomaly}A). Grey lines show 99\% percentile intervals from beta-binomial Monte Carlo simulations.}
\label{fig:othercountries}
\end{figure}

Finally, we applied our analysis to three recent elections outside Russia: one in Spain, one in Germany, and one in Poland. In each case we computed 1000 Monte Carlo iterations and in each case the number of integer polling stations was well inside the 99\% percentile interval of the Monte Carlo values (Figure~\ref{fig:othercountries}), demonstrating that the number of integer values was not at all anomalous. In fact, in each case the number of integer polling stations was very close to the mean Monte Carlo value, demonstrating adequacy of the model.

\subsection{Specific integers}

\begin{figure}[!tph]
\includegraphics[width=0.95\textwidth]{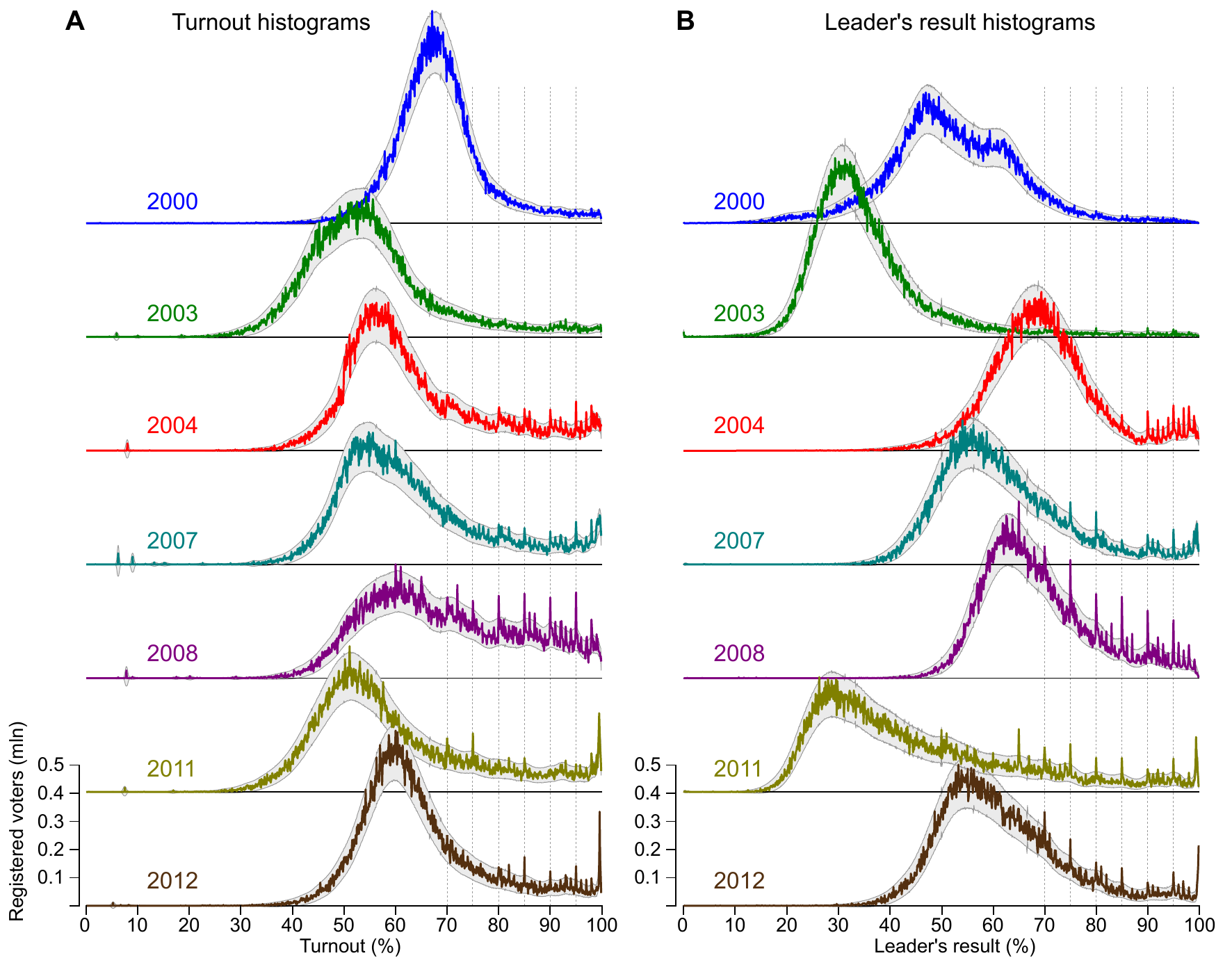}
\caption{\textbf{(A)} Turnout histograms for all elections from 2000 to 2012
(from top to bottom). All histograms show total number of registered
voters at all polling stations with a given turnout ($\pm$0.05\%,
so e.g. the value at 70\% corresponds to turnouts from 69.95\% to
70.05\%). Shaded areas show 99\% percentile intervals of 10~000 respective
Monte Carlo simulations. Values at 100\% turnout are not shown. \textbf{(B)} Histograms of leader's result. Values at 100\%
not shown for consistency with turnout.}
\label{fig:hist}
\end{figure}

Which integer percentages contributed most to the integer anomaly? To answer
this question, we considered histograms of turnout and leader's result
for each year (Figure~\ref{fig:hist}). To account for different sizes of polling
stations, we selected all polling stations exhibiting a particular
turnout or leader's result (in $\pm 0.05$\% bins) and plotted the total number of registered
voters on these polling stations. The same histograms were computed
for the surrogate data obtained with Monte Carlo simulations, and
distributions of these surrogate histograms (99\% percentile intervals)
are shown on Figure~\ref{fig:hist} as gray shaded areas. Note that the Monte Carlo
histograms follow the empirical ones very closely (except for a number of integer peaks, see below), demonstrating self-consistency of the Monte Carlo procedure.

\begin{figure}[!bph]
\includegraphics[width=0.95\textwidth]{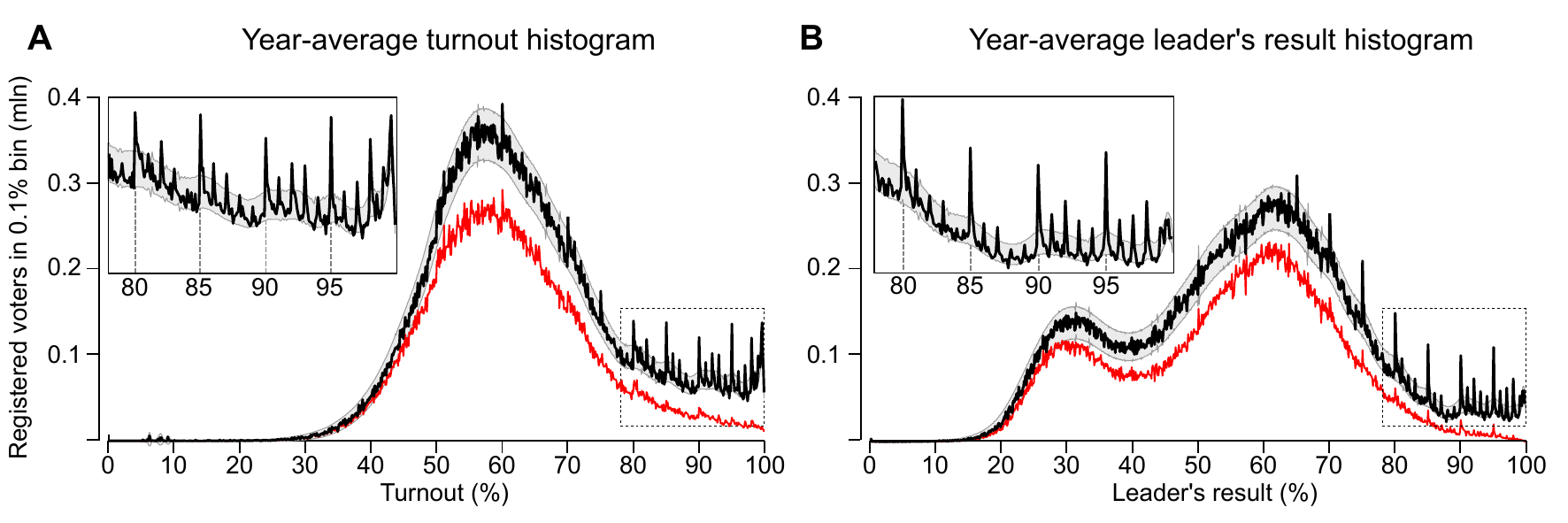}
\caption{\textbf{(A)} Average over all turnout histograms from Figure~\ref{fig:hist}A.
Inset provides a zoom-in overview of the peaks at high percentage
values. Shaded area shows 99\% percentile intervals of the year-averaged
Monte Carlo histograms. Red curve shows the same average histogram
obtained after excluding 15 regions of Russia with particularly pronounced
prevalence of integer polling stations. \textbf{(B)} Average over
leader's result histograms from Figure~\ref{fig:hist}B. }
\label{fig:meanhist}
\end{figure}

Starting from 2004, all empirical histograms exhibit pronounced sharp
peaks at all integer percentage values of turnout and/or leader's result
above $\sim$70\%. At multiples-of-five (75\%, 80\%, 85\%, etc.) percentage
values that are arguably more appealing, the peaks are particularly
high \citep{buzin2008crime, mebane2009comparative}. 
Nevertheless, smaller but denser peaks are also apparent at integer percentage
values above $\sim$80\% (e.g. at 91\%, 92\%, 93\%, etc.). These peaks
often reach well outside of the shaded Monte Carlo area, meaning that for many
of them their individual $p$-values are less than 0.0001. Fourier analysis
confirms that the peaks are strictly equidistant with periods of
1\% and 5\% (Figure~\ref{fig:fourier}) and that such periodic peaks appear only at high percentages, namely above $\sim 70\%$ (Figure~\ref{fig:spectrograms}).

We carefully checked that these peaks are not
the artifacts of division of small integers \citep{johnston1995}. Such artifacts
can be observed in the election histograms if one chooses a very small
bin size and counts polling stations directly instead of weighting
them by registered voter counts (as we do). This allows small polling
stations to contribute strongly to the distributions, leading to the
artifact peaks at fractions with small denominators (such as 1/2,
2/3, 3/4, i.e. 50\%, 66\%, 75\%). The peaks visible on Figure~\ref{fig:hist}
are totally different from such artifacts, because (i) they are strictly
periodic, (ii) they are never observed at 50\% where the artifacts
would be strongest, and (iii) they do not appear in Monte Carlo simulations
that involve exactly the same type of integer divisions\footnote{A convenient way to get rid of such artifacts is to add a random number sampled from a uniform distribution $\mathcal U(-0.5, 0.5)$ to the nominator of each fraction, e.g. to the number of given ballots when computing the turnout. This does not noticeably influence the turnout value, but eliminates the problems associated with the division of integers. We did not apply this procedure here, as the artifacts were negligible.}.

The integer peaks appear at the same positions in all years since
2004, demonstrating that the same integer numbers remain to be particularly
appealing. Due to this fact, averaging the histograms over the years
increases signal-to-noise ratio (Figure~\ref{fig:meanhist}) allowing us to study the
fine structure of the peaks. As can be seen in the insets of Figure~\ref{fig:meanhist},
the peaks are asymmetric: sharp raising left flank is followed by
a relaxed right tail.

\begin{figure}[h]
\centering
\includegraphics[width=0.5\textwidth]{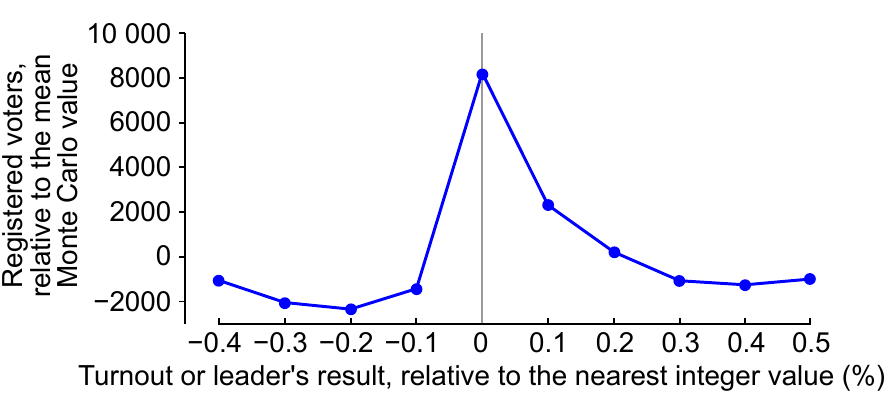}
\caption{Average shape of the integer peaks in Figure~\ref{fig:meanhist}.}
\label{fig:meanpeak}
\end{figure}

To inspect this effect closer, we computed the average shape of all integer peaks in Figure~\ref{fig:meanhist}. We subtracted the respective Monte Carlo mean values from the year-averaged turnout (Figure~\ref{fig:meanhist}A)
and leader's result (Figure~\ref{fig:meanhist}B) histograms, and averaged them over
all 1\%-long intervals around integer values (so the average was computed
over 198 intervals, 99 for turnout and 99 for leader's result, from
1\% to 99\%). Figure~\ref{fig:meanpeak} confirms that the resulting shape is indeed asymmetric. This behavior is consistent
with the interpretation that the polling station officials seem
to be motivated to report a turnout or result which is ``just
above'' an appealing integer value, rather than
``just below'' it. This
leads to depletion of the votes right before an integer value, a peak
at the exact integer value, and subsequent relaxation until the next
integer value comes into play. This peculiar shape explains the decrease of $z$-scores of the main anomaly as the window size around integer percentages gets too broad (Figure~\ref{fig:anomaly}C).

\subsection{Geographic distribution of integer anomaly}

Geographically, polling stations contributing to the integer anomaly are not evenly
distributed across Russia, but tend to cluster in certain regions.
To show this, we computed the turnout and leader's result histograms
for each of the 7 elections in each of the 83 Russian regions separately
and ranked the regions by the magnitude of the most conspicuous integer
peak across years (Figure~\ref{fig:regions} and Table~\ref{tab:regions}).

\begin{figure}[h]
\centering
\includegraphics[width=1\textwidth]{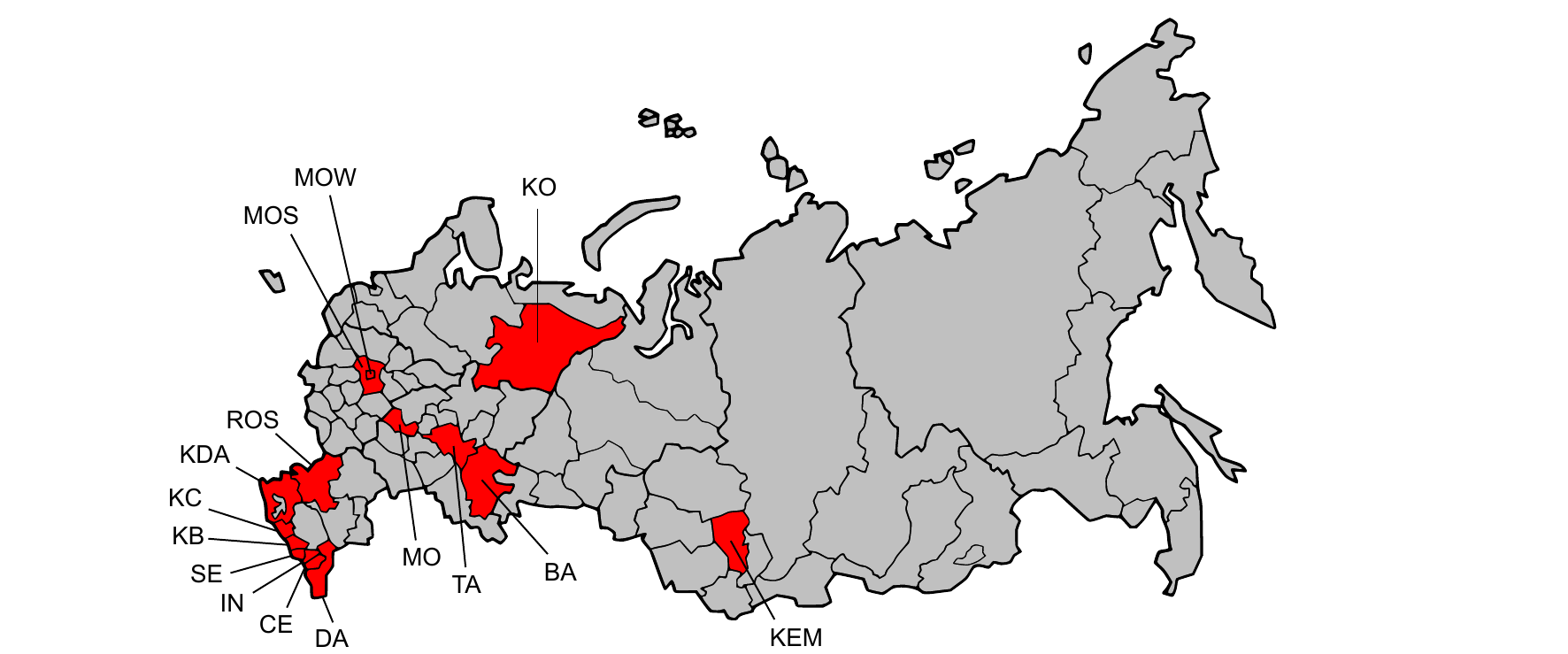}
\caption{Fifteen regions identified as contributing the most to the integer anomaly. Regions are designated by their ISO codes (see Table~\ref{tab:regions}).}
\label{fig:map}
\end{figure}

We found that the vast majority of the integer peaks originated from 15 regions,
with the city of Moscow and the Moscow Region among them (Figure~\ref{fig:map}). If these
15 regions (comprising $\sim$33 mln voters, $\sim$30\% of the national
total) are excluded from the analysis, the integer peaks in both turnout
and leader's result histograms become negligibly small (Figure~\ref{fig:meanhist},
red lines). Geographical clustering of integer polling stations strongly
suggests that there existed tacit inducement, encouragement or even
coordinating directives from the higher electoral commissions at the
region level towards the individual polling stations (note that each
region in Russia has its own electoral commission). Such conduct was rationalized by \citet{kalinin2010understanding} as regions signaling their loyalty to the center.

\subsection{Relation to other electoral anomalies}

In a recent study \citep{klimek2012}, two features of post-2004 Russian
elections have been suggested as potential falsification fingerprints:
 high correlation between turnout and leader's result,
and high amount of polling stations with both turnout
and leader's result close to 100\%. When the 15 aforementioned regions
are excluded from the analysis, both features substantially weaken
or disappear entirely.

\begin{figure}
\includegraphics[width=1\textwidth]{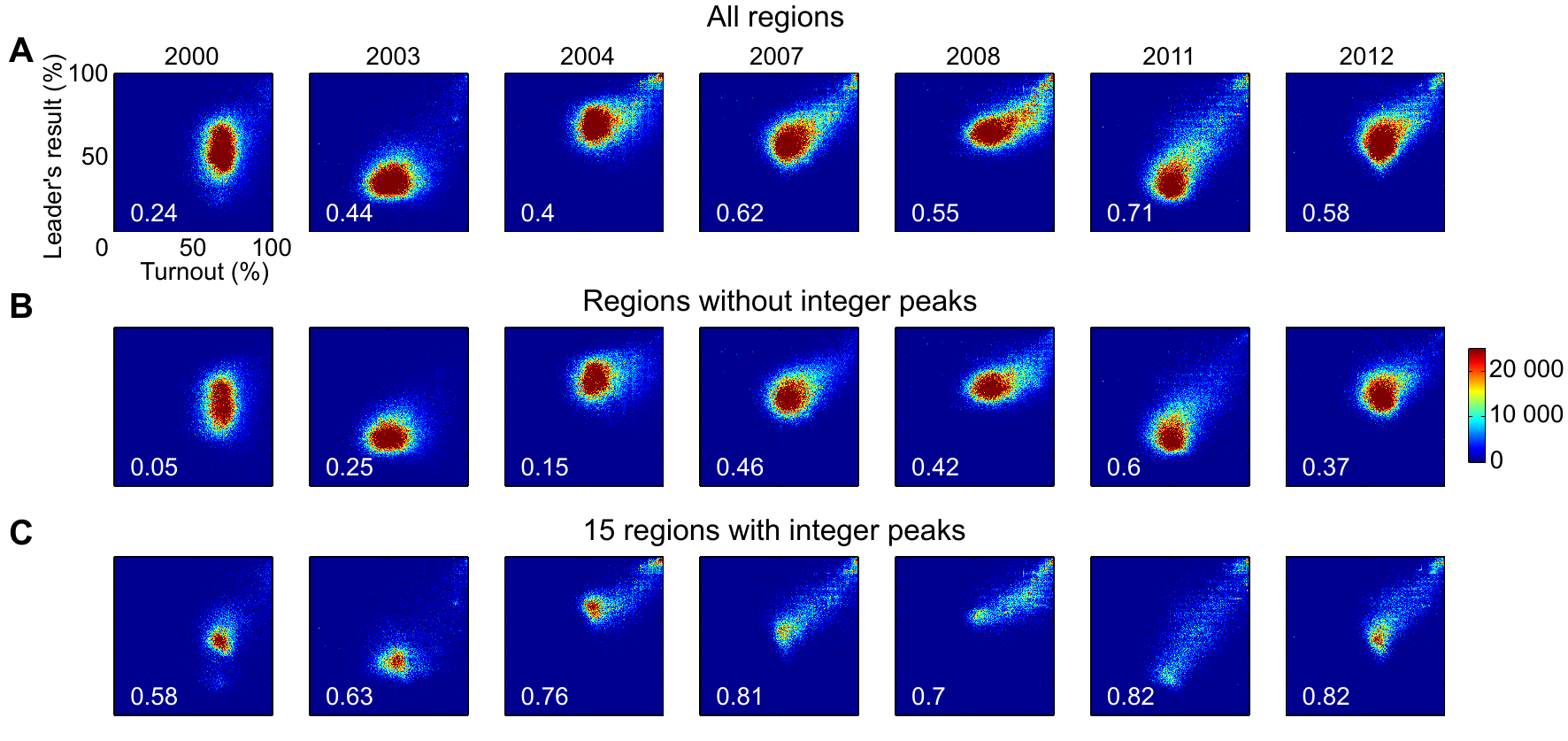}
\caption{\textbf{(A)} 2D histograms for all years: horizontal axis shows turnout
in 0.5\% bins, vertical axis shows leader's result in 0.5\% bins,
number of voters in the respective polling stations is colour-coded. 
\textbf{(B)} The same for all regions apart from 15 regions demonstrating
most prominent integer anomalies. 
\textbf{(C)} The same only for 15
regions demonstrating most prominent integer anomalies. Summing the histograms
on panels (B) and (C) gives exactly the histograms from panel (A).
Pearson correlation coefficient between turnout and leader's
result (across all polling stations) is shown in the lower left corner
of each diagram.}
\label{fig:klimek}
\end{figure}

This is illustrated by 2D histograms similar to the ones  used in  \citep{klimek2012} (Figure~\ref{fig:klimek}). Klimek et al. hypothesized that there are two main types of falsifications: ``incremental fraud'' when some extra ballots for the leader are added (ballot stuffing), and ``extreme fraud''
when a polling station reports almost 100\% turnout and almost 100\%
leader's result. On a 2D turnout-result histogram the first type of
fraud shows as an extremely high correlation between turnout and leader's
result, while the second type of fraud gives rise to a separate second
cluster near 100\%-turnout, 100\%-result point. Indeed,
both features are present in Russian elections after 2004 (Figure~\ref{fig:klimek}A).

When the 15 regions demonstrating most prominent integer anomalies are excluded, the high-percentage cluster 
fully vanishes, and the correlation between turnout and leader's result substantially weakens (Figure~\ref{fig:klimek}B). On the other hand, if only these 15 regions are used for the histograms, 
both anomalies become very prominent (Figure~\ref{fig:klimek}C). 

The fact that
the regions with the highest level of integer outcome anomaly are
almost exactly those exhibiting other suspicious features, provides
justification to the previous forensic methods \citep{mikhailov2004, myagkov2009, klimek2012, simpser2013}
and lends additional support to the current interpretation.

\section{Conclusions}

In sum, our results present a historical overview of
the 2000--2012 Russian elections  based on a novel
statistical fraud indicator. The elections in 2000 and in 2003 do not appear to show any strong statistical anomalies. The anomalous integer-value peaks indicative of electoral manipulations popped up in 2004 and have persisted in the election data ever since, reaching a  maximum in 2008 elections won by Dmitry Medvedev. What exactly happened during the three months between December 2003 and March 2004 when the respective elections were held, is an interesting politological question which however falls outside of the scope of the current paper. It remains to be seen if the anomalies discussed in
this paper will show up in the upcoming 2016 parliament elections.

One of the limitations of the forensic method presented here is that
it does not provide a way to estimate the overall impact of falsifications:
not all ballots at dishonest polling stations are necessarily fraudulent,
and not all dishonest polling stations report integer percentages.
Nonetheless, agreement of our findings with the previous studies \citep{klimek2012}
at the level of regions makes us believe that the excess of integer percentages is just a tip-of-the-iceberg effect unforeseen by
the forgers. The real significance of the fraud indicator described
herein is in its irrefutable character.

In a wider perspective, the methodology developed in this paper can also be useful for forensic studies of any datasets where percentages, or
fractions, are of particular interest. Apart from the electoral data, this might be the case for
scientific datasets where our method can possibly inform future investigations of scientific misconduct \citep{simonsohn2013}.

\section*{Acknowledgements}
We thank Sergey Slyusarev, Boris Ovchinnikov, Peter Klimek, and Uri Simonsohn for comments
and suggestions, Alexey Shipilev for providing 2011--2012 election data
on the fly, Alexander Shen for enlightening comments on statistical testing, and G\"{u}nter Ziegler for providing us with the raw election data from Germany.

\begin{supplement}[id=suppA]
\stitle{Election data}
\slink[doi]{https://dx.doi.org/10.6084/m9.figshare.3126883}
\sdatatype{.zip}
\sdescription{Datasets used in this study (in tab-delimited plain text format).}
\end{supplement}

\newpage
\bibliography{fingerprints}

\appendix

\setcounter{figure}{0}
\renewcommand{\theHfigure}{\arabic{section}.\arabic{figure}} 
\makeatletter 
\renewcommand{\thefigure}{S\@arabic\c@figure}
\renewcommand{\thetable}{S\@arabic\c@table}
\makeatother

\section{Supplementary figures}\label{app1}

See next page.

\begin{figure}[p]
\includegraphics[width=1\textwidth]{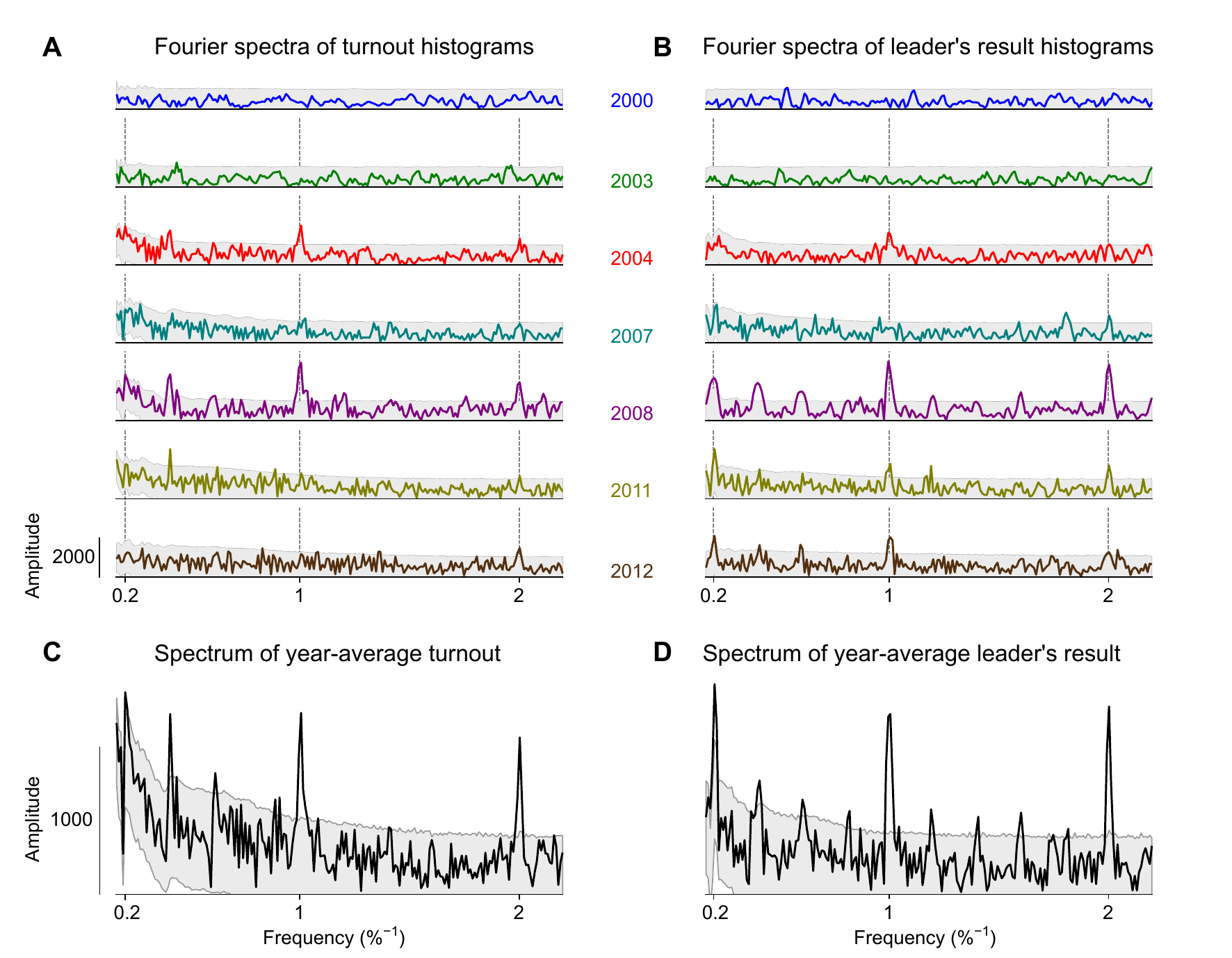}

\caption{\textbf{(A)} Fourier amplitude spectra of turnout histograms from
Figure~\ref{fig:hist}A for all elections from 2000 to 2012 (top to bottom). Harmonics
at 1\%$^{-1}$ and 2\%$^{-1}$ correspond to periodic peaks in Figure~\ref{fig:hist}A
appearing with 1\% intervals, while harmonics at 0.2\%$^{-1}$, 0.4\%$^{-1}$
etc. are characteristic for periodic peaks appearing every 5\%. \textbf{(B)}
Fourier spectra of leader's result histograms from Figure~1B. \textbf{(C)}
Fourier spectrum of the year-averaged turnout histogram from Figure~\ref{fig:meanhist}A. Note that as the peaks become more prominent
in the year-average histograms, the corresponding
peaks in the Fourier spectrum are also boosted.
\textbf{(D)} Fourier spectrum of the year-averaged leader's result
histogram from Figure~\ref{fig:meanhist}B. Shaded areas on all panels show 99\% percentile intervals
of the respective Monte Carlo spectra. The Fourier amplitude spectra
were computed as absolute value of the discrete Fourier transform
normalized by the sampling length ($100/0.1=1000$). }
\label{fig:fourier}
\end{figure}

\begin{figure}
\includegraphics[width=1\textwidth]{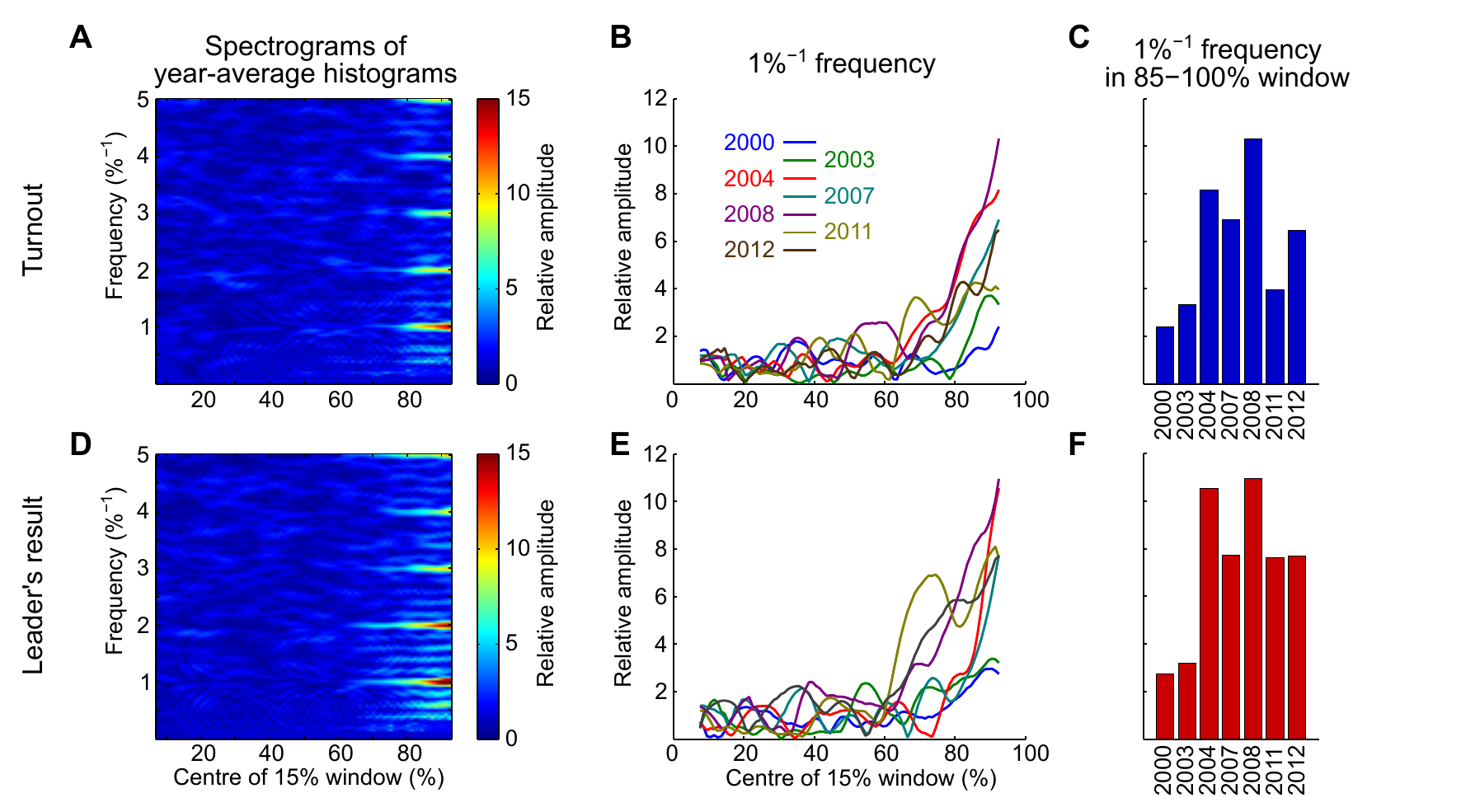}

\caption{(A) Fourier spectrogram of the year-averaged turnout histogram from
Figure~\ref{fig:meanhist}A. The Fourier transform was computed in a sliding 15\%-wide
Hamming window. The horizontal axis shows the position of the centre
of the window and ranges from 7.5\% to 92.5\%. The vertical axis shows
the frequency and ranges from 0 to 5\%$^{-1}$ (with 5\%$^{-1}$ being
the Nyquist frequency given our resolution of 0.1\%). The spectrogram
was normalized (separately for each percent-frequency value) by the
average over 10\,000 spectrograms obtained with year-averaged Monte
Carlo histograms (see Figure~\ref{fig:meanhist}A). Resulting values are colour-coded.
\textbf{(B)} The same procedure was repeated for each year separately
using histograms from Figure~\ref{fig:hist}A, and the relative amplitude of 1\%$^{-1}$
harmonic (representing amplitudes of both 5\% and 1\% peaks) is shown
for each year. The interpretation of this panel is that the periodic peaks begin to appear around high values
of the turnout and result ($\sim$70\%), and the magnitude of peak harmonics
steadily increases all the way up
to 100\%.\textbf{(C)} The relative amplitude of 1\%$^{-1}$
harmonic in the last 85--100\% window for each year. The values correspond
to the rightmost values of the functions displayed on Panel (B).\textbf{
(D--F)} The same for leader's result histograms. }
\label{fig:spectrograms}
\end{figure}

\begin{figure}
\includegraphics[width=1\textwidth]{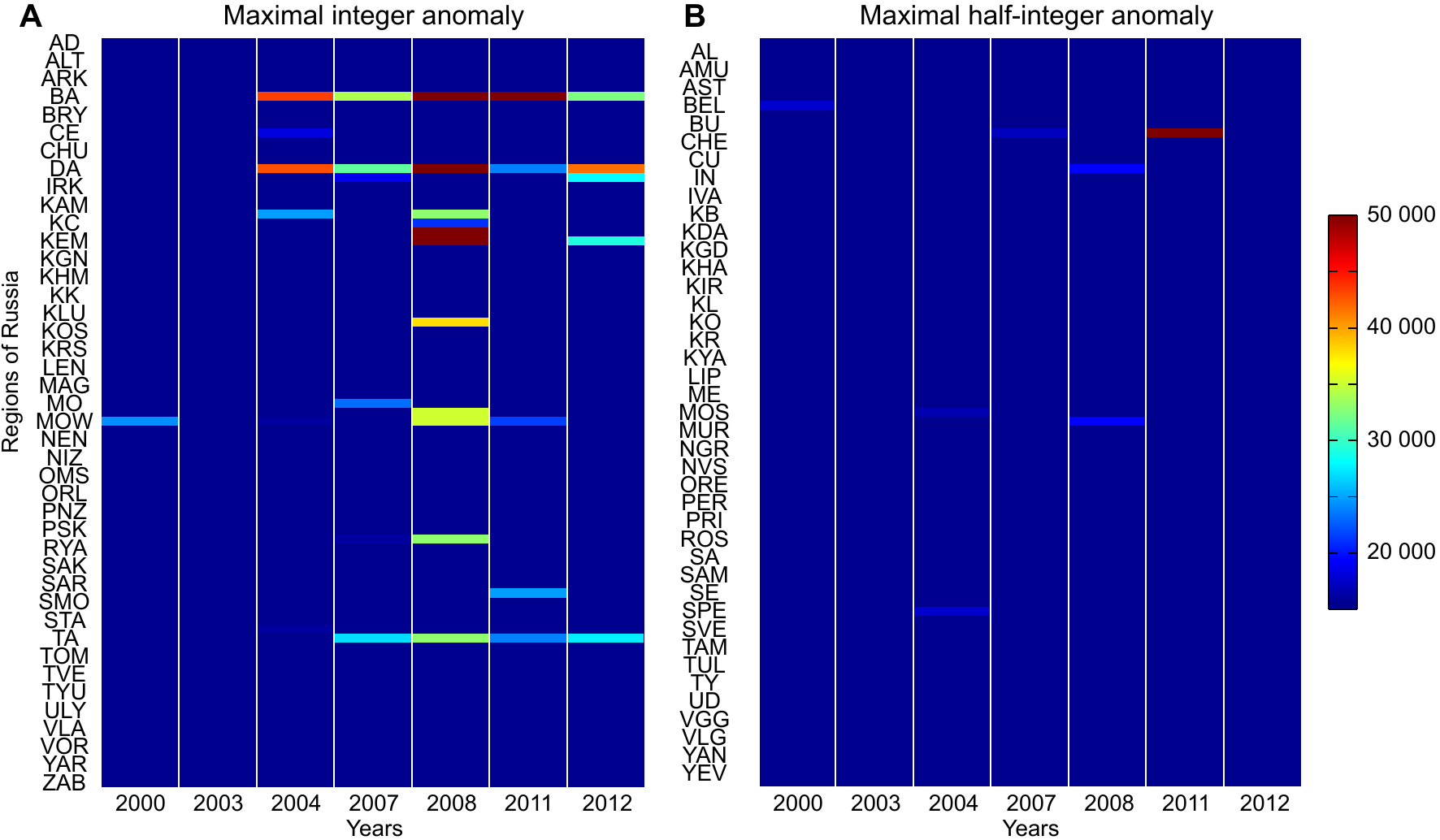}
\caption{\textbf{(A)} Amplitude of the most prominent integer peak for each
year (horizontal axis) in each of the 83 regions (vertical axis).
The amplitude was defined as the difference between an empirical value
and a corresponding mean Monte Carlo value. The most prominent integer
peak was identified as the one having maximal amplitude over all integer
values between 70\% and 99\% in both turnout and leader's result histograms
(i.e. the maximum over $29\cdot2=58$ values). There are 15 regions
(see Table~\ref{tab:regions}) exhibiting noticeable integer peaks, many of them in
several elections. \textbf{(B)} For comparison: amplitude of the most
prominent peak over all half-integer percentage values between 70.5\%
and 99.5\%. These data show that there are much fewer peaks located
at half-integer positions (apart from the one in Republic of Chechnya
in 2011 located at 99.5\% and corresponding to 99.5\% result for Vladimir
Putin at the polling stations in this region). Regions are marked
with their ISO 3166-2 codes, with RU- prefix omitted. Half of the regions are named on the left and the other half in the middle. Both panels contain the same number of rows (regions).}
\label{fig:regions}
\end{figure}

\begin{figure}
\renewcommand\figurename{Table}     

\begin{tabular}{llc}
\textbf{ISO code} & \textbf{Region name} & \textbf{Maximal integer anomaly ($10^{3}$) }\tabularnewline
\hline 
\noalign{\vskip0.2cm}
DA & Dagestan, Respublika  & 87\tabularnewline
BA & Bashkortostan, Respublika & 64\tabularnewline
KEM & Kemerovskaya Oblast  & 52\tabularnewline
KDA & Krasnodarskiy Krai  & 51\tabularnewline
KO & Komi, Respublika  & 38\tabularnewline
MOS & Moskovskaya Oblast  & 35\tabularnewline
MOW & Moscow & 34\tabularnewline
KB & Kabardino-Balkarskaya Respublika  & 33\tabularnewline
TA & Tatarstan, Respublika  & 33\tabularnewline
ROS & Rostovskaya Oblast  & 33\tabularnewline
IN & Ingushetiya, Respublika  & 28\tabularnewline
SE & Severanaya Osetiya-Alaniya, Respublika & 24\tabularnewline
MO & Mordoviya, Respublika  & 23\tabularnewline
KC & Karachayevo-Cherkesskaya Respublika & 21\tabularnewline
CE & Chechenskaya Respublika  & 18\tabularnewline
\end{tabular}
\caption{Top 15 regions contributing to the integer anomaly. For each region
we report the height of the maximal integer peak, where maximum is
taken over all years, over both turnout and leader's result, and over
all integer percentage values from 70\% to 99\%. Peak heights are
measured relative to the mean Monte Carlo value. ISO codes are given
according to the ISO 3166-2 standard, with RU- prefix omitted.}
\label{tab:regions}
\end{figure}

\end{document}